\documentclass[aps,prb,twocolumn,showpacs,citeautoscript,groupedaddress]{revtex4}
\usepackage{graphicx}
\usepackage{SIunits}
\usepackage{color}

\usepackage{ifpdf}
\ifpdf
\fi

\def\figref#1{Fig.~\ref{#1}}
\def\red#1{{\color{black}#1}}

\def\twosome#1{%
  \begin{minipage}[b]{.48\textwidth}\centering%
  \includegraphics{fig#1a}\\{}(a)\end{minipage}
  \begin{minipage}[b]{.48\textwidth}\centering%
  \includegraphics{fig#1b}\\{}(b)\end{minipage}}

\def\addfig#1{\includegraphics[width=.42\textwidth]{#1}}

\begin{document}

\title{Creating double negative index materials using the Babinet principle with one metasurface}

\author{
\textbf{Lei Zhang}$^{1,*}$
\textbf{Thomas Koschny}$^{1}$,
\textbf{and C.\ M.\ Soukoulis}$^{1,2,\dag}$\\
\small\itshape $^{1)}$Ames Laboratory---U.S.~DOE, and Department of Physics and Astronomy,\\
\small\itshape Iowa State University, Ames, Iowa 50011, USA\\
\small\itshape $^{2)}$Institute of Electronic Structure and Lasers (IESL),\\
\small\itshape FORTH, 71110 Heraklion, Crete, Greece
}

\date{\today}

\begin{abstract}
Metamaterials are patterned metallic structures which permit access to a novel
electromagnetic response, negative index of refraction, impossible to achieve
with naturally occurring materials. Using the Babinet principle, the
complementary split ring resonator (SRR) is etched in a metallic plate to
provide negative $\epsilon$, with perpendicular direction. Here we propose a
new design, etched in a metallic plate to provide negative magnetic
permeability, $\mu$, with perpendicular direction. The combined
electromagnetic response of this planar metamaterial, where the negative $\mu$
comes from the aperture and the negative $\epsilon$ from the remainder of the
continuous metallic plate, allows achievement of a double negative index metamaterial
(NIM) with only one metasurface and strong transmission. These designs can be
used to fabricate NIMs at microwave and optical wavelengths and three
dimensional metamaterials.
\end{abstract}

\pacs{78.67.Pt, 78.20.-e,  42.25.Bs}

\maketitle

\section{Introduction} Metamaterials are artificial materials that can be
engineered to exhibit fascinating electromagnetic properties that do not occur
in nature, such as negative refractive index, “perfect” imaging and
electromagnetic cloaking \cite{Shalaev2007, Soukoulis2007, Zheludev2010,
Soukoulis2010, Boltasseva2011, Soukoulis2011,Liu2011,Tassin2012,Schurig2006},
Chiral metamaterials have shown giant optical activity \cite{Rogacheva2006},
circular dichroism \cite{Gansel2009}, negative refraction \cite{Pendry2004,
Plum2009, Zhou2009, Zhang2009}, and possible reversal of the Casimir force
\cite{Zhao2009}. Most of the applications critically required an negative
index of refraction in three dimensions (3D). It is an open question, whether
there is any 3D metamaterial design working for optical frequencies still
feasible to fabricate. Here, we theoretically proposed a new design that
provides negative index of refraction with one surface and can be used to
fabricate 3D metamaterials. This is an important step towards the
realization of bulk 3D \red{ double negative index metamaterials (NIMs)} at optical wavelengths. Previous theoretical
designs \cite{Soukoulis2011} obtain NIM in 3D at GHz and THz. Two groups
independently proposed fully isotropic bulk magnetic metamaterial designs,
based on SRRs arranged in a cubic lattice by employing spatial symmetries
\cite{Padilla2007,Baena2007}. On the other hand, some designs of 3D isotropic
NIMs exist, but fabricating them has remained a challenging task and virtually
impossible at optical frequencies. For example, Koschny et al.
\cite{Koschny2005}, designed an early example of an isotropic NIM. However,
high-constant dielectric assumed across the gaps of the SRRs rendered the
experimental realization impractical. Alternative approaches have also been
investigated, including direct lasing fabrication
\cite{Soukoulis2011,Rill2008,Guney2010}, and high refractive index spheres
\cite{Soukoulis2011,ZhaoQ2009}.

Double negative index materials exhibit simultaneously negative magnetic
permeability, $\mu$, and electric permittivity, $\epsilon$, over a common
frequency range \cite{Veselago}. Negative permeabilities are the result of a
strong resonant response to an external magnetic field; negative permittivity
can appear by either a plasmonic or a resonant response (or both) to an
external electric field. Pendry et al. \cite{Pendry} suggested a double
metallic SRR design for negative $\mu$ and a parallel metallic wire periodic
structure to give negative $\epsilon$. Several variations of the initial
design have been studied \cite{Zhou,Shalaev2007,Soukoulis2007, Soukoulis2011}.
For example, a single ring resonator with several cuts has been proven capable
of reaching negative $\mu$ at a higher frequency \cite{Zhou}, and cut wire
pairs and fishnet structures \cite{Shalaev2007,Soukoulis2007, Soukoulis2011}
allow for a negative magnetic response at optical frequencies. In these old
designs, it needs two particles, one particle gives negative $\mu$ and the
other gives negative $\epsilon$ at the same frequency range. Later, Rockstuhl and Lederer proposed a new design\cite{Rockstuhl2}, nanoapertures embedded in a continuous metal film, that gives $\epsilon<0$ from the metallic surface and a resonant $\mu$, which, however, did not reach $\mu<0$ because of the weak resonance. Since the imaginary part of $\mu$ is large, $\mu>0$ and $\epsilon<0$ will \red{still} give negative index of refraction\cite{Rockstuhl2, Zhou2006}. \red{This design suffers from large losses, partially due to the weak magnetic dipole of the resonance but also - like other resonant structures at optical wavelengths - due to the generally high ohmic losses in metals at high frequencies}\cite{Soukoulis2010}. It\ensuremath{'}s very
interesting to develop a new design that will use only one ``particle'' on one
surface that will provide negative index of refraction and also $\mu<0$ and
$\epsilon<0$, with parallel direction on the surface. We will use Babinet\ensuremath{'}s
principle \cite{Falcone,Zentgrat,Rockstuhl,Chen} to obtain a complementary
structure in a metallic surface that will give $\mu<0$ for the first time,
instead of $\epsilon < 0$ resulting from Babinet-complementary resonators
studied in the previous literature, with perpendicular direction on the
surface. In addition, the remaining continuous metallic surface will give
$\epsilon<0$.

\begin{figure*}[htb]
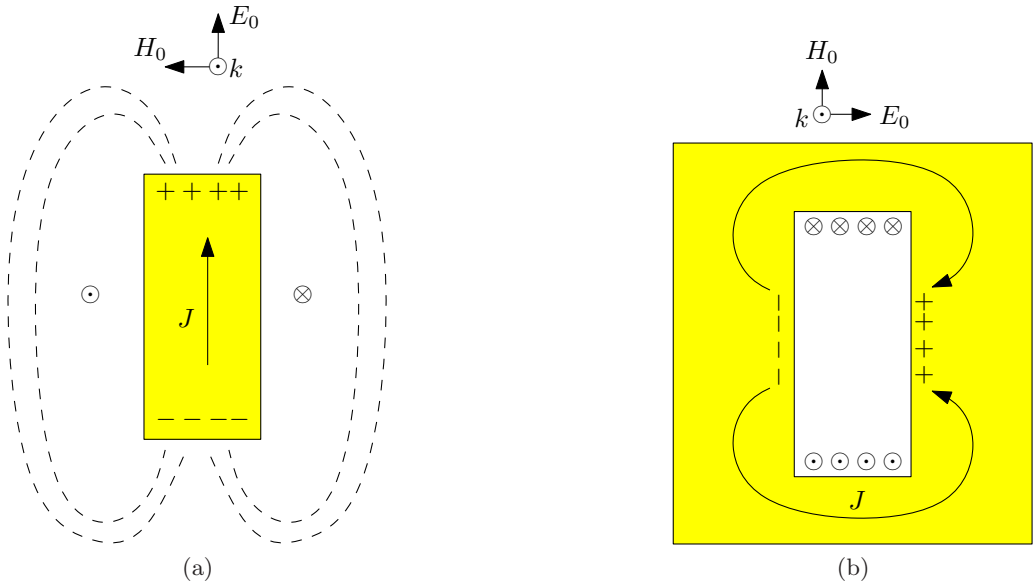

\twosome 1
\caption{(a) Rectangular thin metallic element with the distribution of surface charge and magnetic flux. This design has a strong electric dipole
in the y-direction. (b) The complementary structure of \figref{fig1}a, with the
distribution of the surface charge and the magnetic flux. This complementary
design has a strong electric dipole in the x-direction.}
\label{fig1}
\end{figure*}

\begin{figure*}[htb]
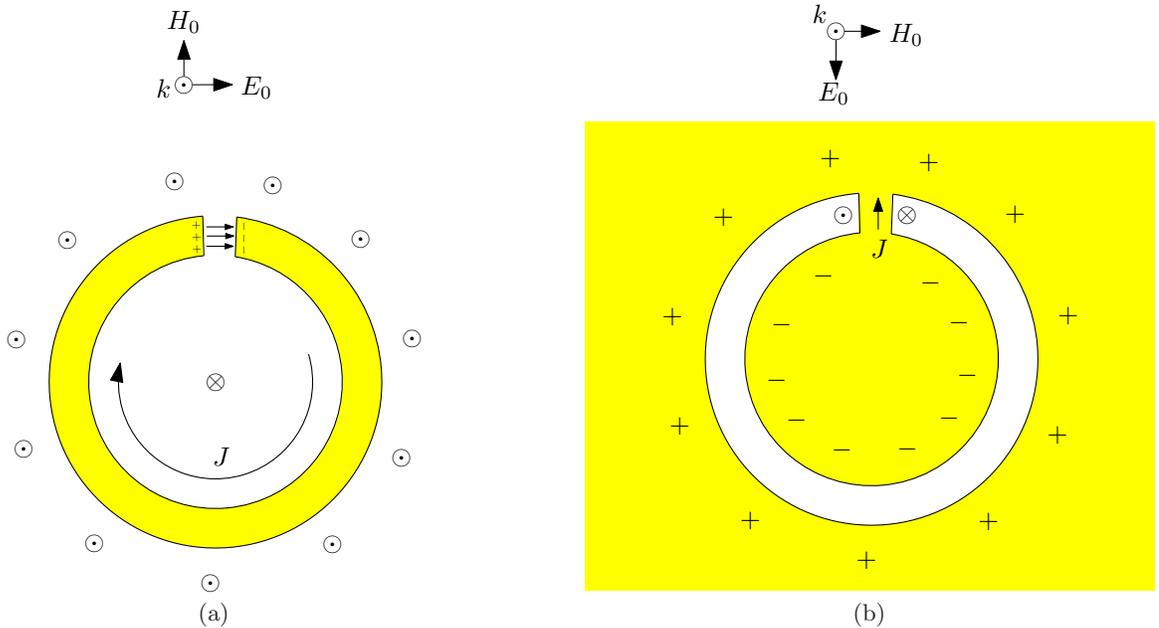

\twosome 2
\caption{Thin metallic SRR structure (a) and the complementary SRR structure
(b) with the distribution of the surface charges and the magnetic flux. The
SRR structure has a strong magnetic dipole perpendicular to the metallic
surface and a minor electric dipole from the SRR gap. The CSRR structure (b)
has a minor electric dipole from the neck of CSRR. }
\label{fig2}
\end{figure*}

Babinet\ensuremath{'}s principle has been used
\cite{Falcone,Zentgrat,Rockstuhl,Chen,Al-Naid,Singh,Bitzer} to design
artificial planar metamaterials, which can provide only negative electric
permittivity. Such inverse elements, like complementary SRRs (CSRRs), have
been proposed as an alternative to conventional metallic wires for the design
of planar metamaterials, which give $\epsilon<0$. Complementary metamaterials
show \cite{Chen,Bitzer} similar properties as their inverse structures. The
transmission coefficient, $t^c$, for the CSRR is related to the transmission
coefficient, $t$, for the SRR by $t+t^c=1$. In addition, according to
Babinet\ensuremath{'}s principle, their transmission and reflection behavior, as well as
their scattered electric and magnetic fields, are interchanged
\cite{Falcone,Zentgrat,Rockstuhl,Chen,Al-Naid,Singh,Bitzer}. These
complementary designs provide an effective negative permittivity,
\cite{Falcone, Zentgrat, Rockstuhl, Chen} with perpendicular direction. In
this manuscript, we will use Babinet\ensuremath{'}s principle to design complementary
structures that will provide an effective negative permeability and the
metallic surface to give negative permittivity. Therefore, we will obtain
negative $n$ with only one planar metamaterial, with parallel direction of the
surface of planar metamaterial.

\begin{figure*}[tb]
\twosome 3
\caption{Thin metallic design (a) and the complementary design (b) with the
distribution of the surface charge and the magnetic flux. The design (a) has
an electric dipole and the complementary design (b) has a residual magnetic dipole. }
\label{fig3}
\end{figure*}

\section{Babinet\ensuremath{'}s principle}

Babinet\ensuremath{'}s principle relates the fields scattered by two complementary planar
structures of arbitrary shape made of infinitely thin, perfectly conducting
sheets. If we consider an incident electromagnetic (EM) field, $E_0, B_0$,
then its complementary EM field, ${E_0}^c, {B_0}^c$, is defined \cite{Falcone,
Jackson} as ${E_0}^c=-c{B_0}^c$ and ${B_0}^c={E_0}^c/c$, which corresponds to
a \unit{90}{\degree} rotation around the propagation axis. If we have a thin
metallic surface, the reflected ($E_r, H_r$) and the transmitted ($E_t, H_t$)
electric and magnetic fields satisfy the following conditions. If we have
charge on the thin metallic surface, the perpendicular reflected electric
field is opposite to the transmitted electric field, i.e., $\vec{n} \cdot
\vec{E_r} = - \vec{n} \cdot \vec{E_t}$, where $\vec{n}$ is the axis vector
perpendicular to the metallic surface. The parallel electric field is
continuous to both sides of the thin metallic surface, i.e., $\vec{n} \times
\vec{E_r}=\vec{n} \times \vec{E_t}$. If the incident EM wave generates current
in the thin metallic surface, the perpendicular magnetic field is continuous,
i.e., $\vec{n} \cdot \vec{H_r} = \vec{n} \cdot \vec{H_t}$ and the parallel
magnetic field is opposite to the two sides of the metallic surface, i.e.,
$\vec{n} \times \vec{H_r}=-\vec{n} \times \vec{H_t}$. If we have a perfectly
conducting surface, no magnetic dipole occurs in the plane due to this
condition $\vec{n} \times \vec{H_r}=-\vec{n} \times \vec{H_t}$. In addition,
no electric dipole occurs perpendicular to the plane, due to the condition
$\vec{n} \cdot \vec{E_r}=-\vec{n} \cdot \vec{E_t}$. However, in-plane electric
dipoles and magnetic
 dipoles perpendicular to the plane are possible. Due to the Babinet principle, we have
two related complementary plane structures. The surface charge in one
structure is related to the magnetic flux in the complement and the opposite
occurs. The magnetic flux is related to the surface charge. Also, the parallel
electric field in the plane can be generated, surface electric current
perpendicular to the electric field (see examples below). In particular, this
means a resonance in one structure is related to a corresponding resonance in
the dual fields in the complementary structure. In \figref{fig1}a, we have a
thin metallic rectangular patch. The propagation direction of the EM field is
perpendicular to the metallic patch and the electric field, $\vec{E_0}$, is
parallel to the long side of the rectangular patch. We have created an
electric dipole in the y-direction of \figref{fig1}a. The electric current,
$\vec{J}$, is along the direction of $\vec{E_0}$ and the magnetic flux is
shown in \figref{fig1}a. In \figref{fig1}b, we presented the
Babinet-complementary structure, where we have a thin metallic surface and a
rectangular hole. As discussed above, the surface charge shown in
\figref{fig1}a can be transformed to the magnetic flux. The positive
(negative) charge, shown in \figref{fig1}a, can be related to the magnetic
flux going inside (outside) the metallic surface (see \figref{fig1}b). The
magnetic flux of \figref{fig1}a can be related with the charge of
\figref{fig1}b. Another example, shown in \figref{fig2}a, is the metallic SRR.
We present the current, the charge and the magnetic flux. In \figref{fig2}a,
we have a strong magnetic dipole along the propagation direction of the EM
wave. There is also a minor electric dipole in the SRR gap. This dipole moment
is responsible for the bianisotropy of the SRR. In \figref{fig2}b, we present
the complementary structure, and used the idea of exchanging surface charge
and magnetic flux. We do not have a magnetic dipole in the complementary SRR
(CSRR), but there is an electric dipole, due to the neck of CSRR, parallel to
the neck.

\begin{figure*}[htb]
\twosome 4
\caption{Thin metallic ``octopus'' design (a) and the complementary design (b)
with the distribution of the surface charge and the magnetic flux. The design
(a) does not have an electric dipole, but an electric quadrupole, and the
complementary design (b) has a magnetic dipole perpendicular to the plane.
Here the propagation direction is parallel to the surface.}
\label{fig4}
\end{figure*}

\begin{figure*}[htb]
\addfig{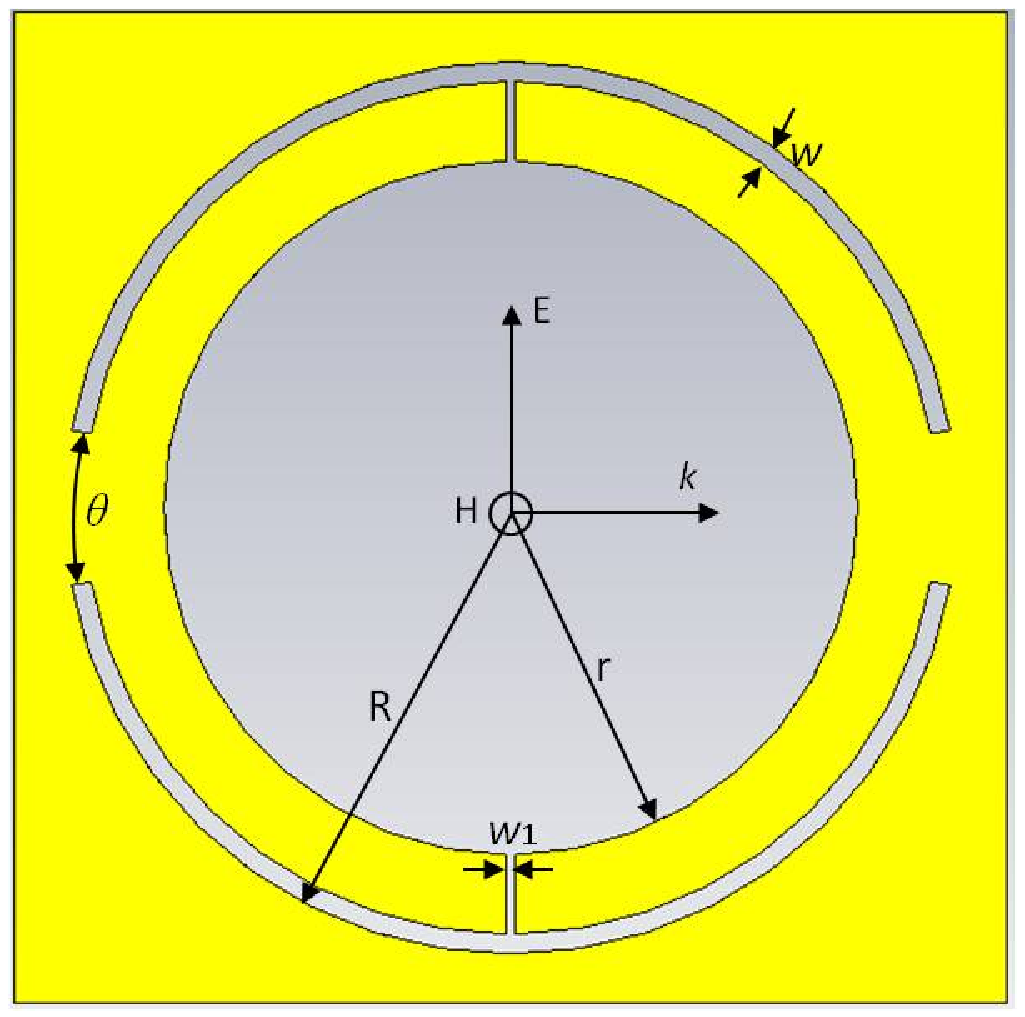}\hspace*{.6in}
\addfig{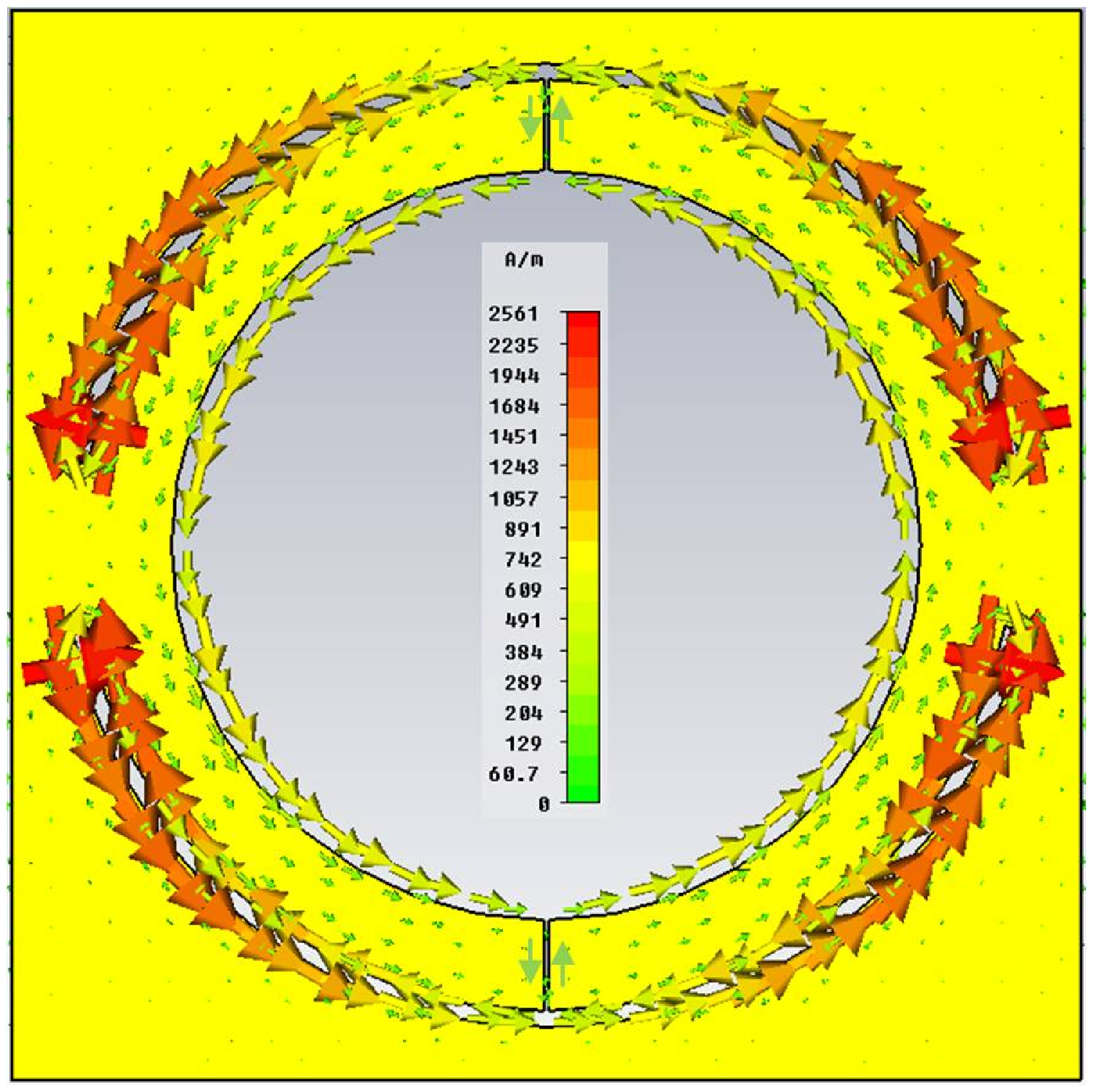}\\(a)\hspace*{3.6in}(b)\\[10pt]
\caption{(a) The complementary design in a thin metallic surface, the
dimensions are $R = 4.5mm$, $ r = 3.5mm$, $w = 0.15mm$, $w_1 = 0.05mm$,
$\theta = 20\degree$. (b) The current density of the complementary design at
9.9 GHz and gives a magnetic dipole perpendicular to the metallic surface. The
propagation direction is parallel to the metallic surface.}
\label{fig5}
\end{figure*}

\begin{figure*}[htb]
\centering
\addfig{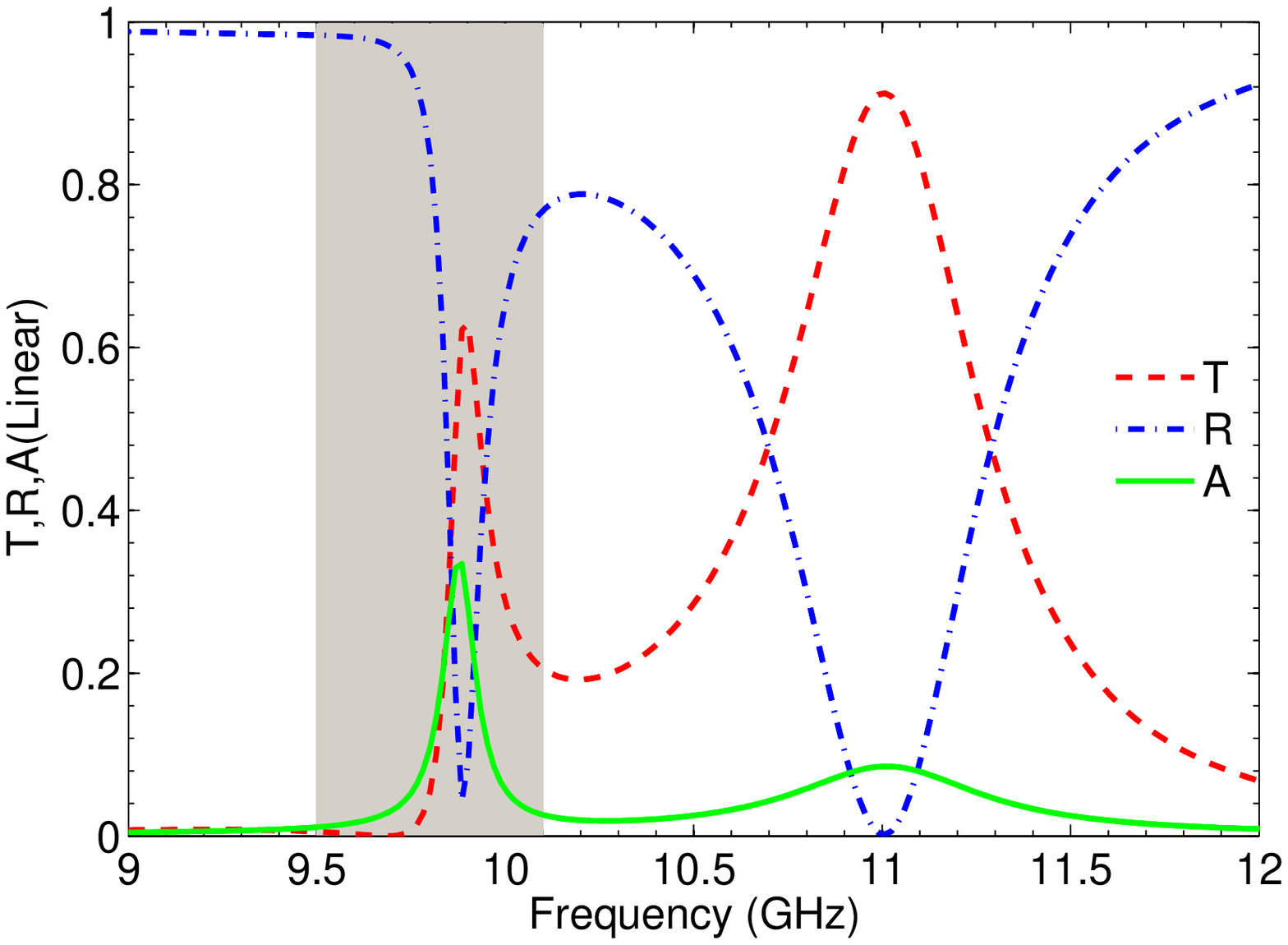}\hspace*{.1in}
\addfig{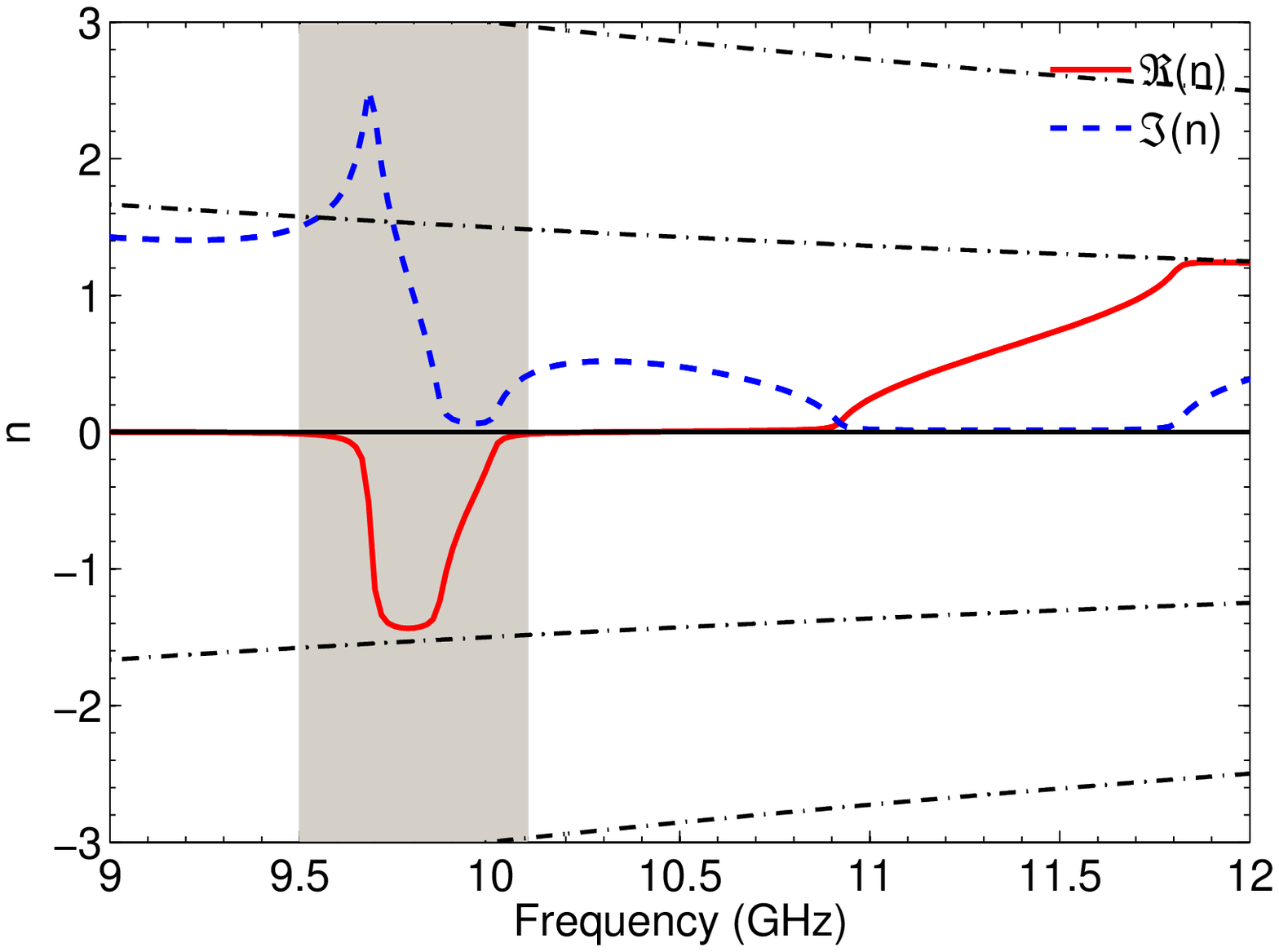}\\(a)\hspace*{3in}(b)\\[3pt]
\addfig{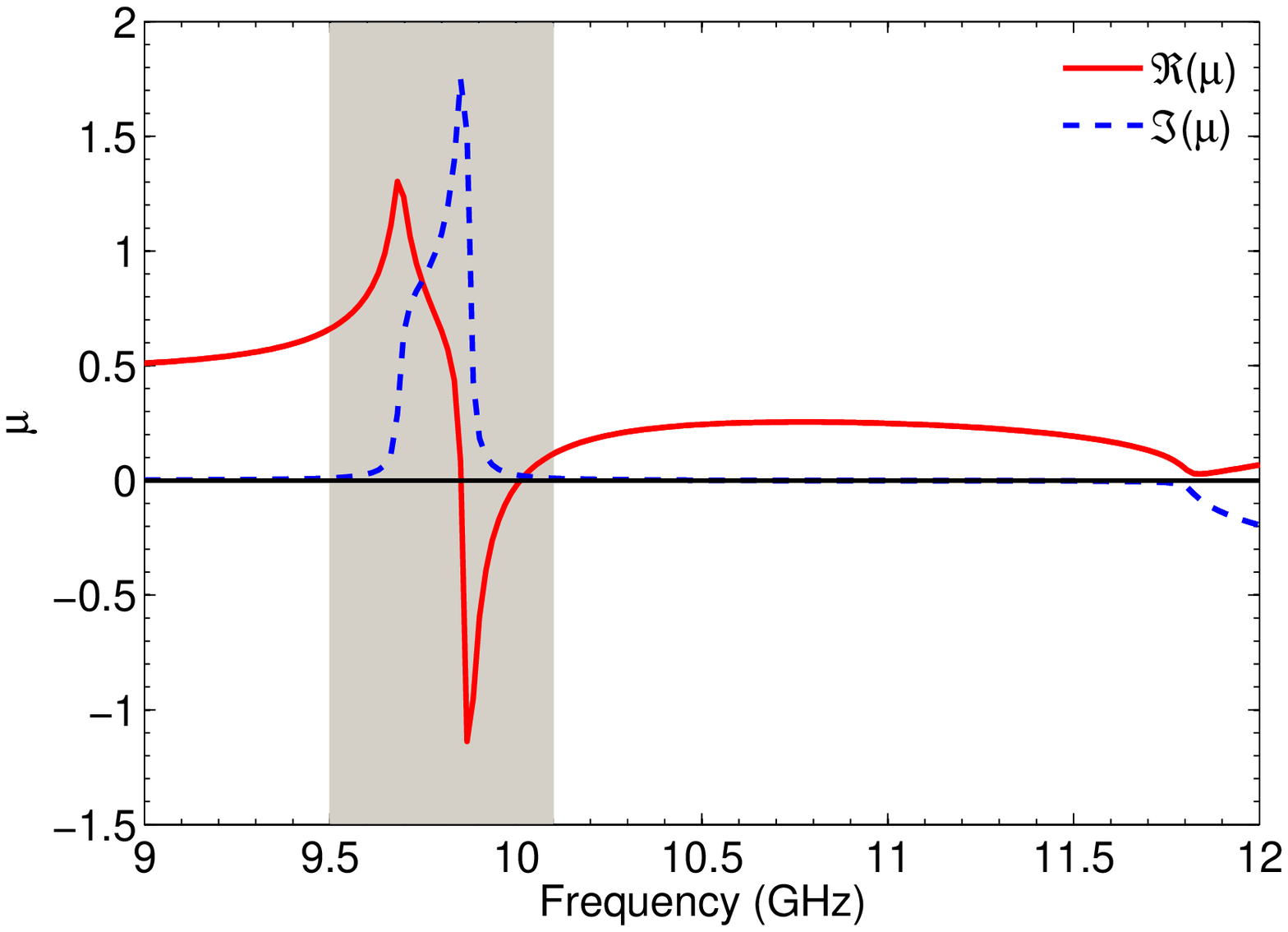}\hspace*{.1in}
\addfig{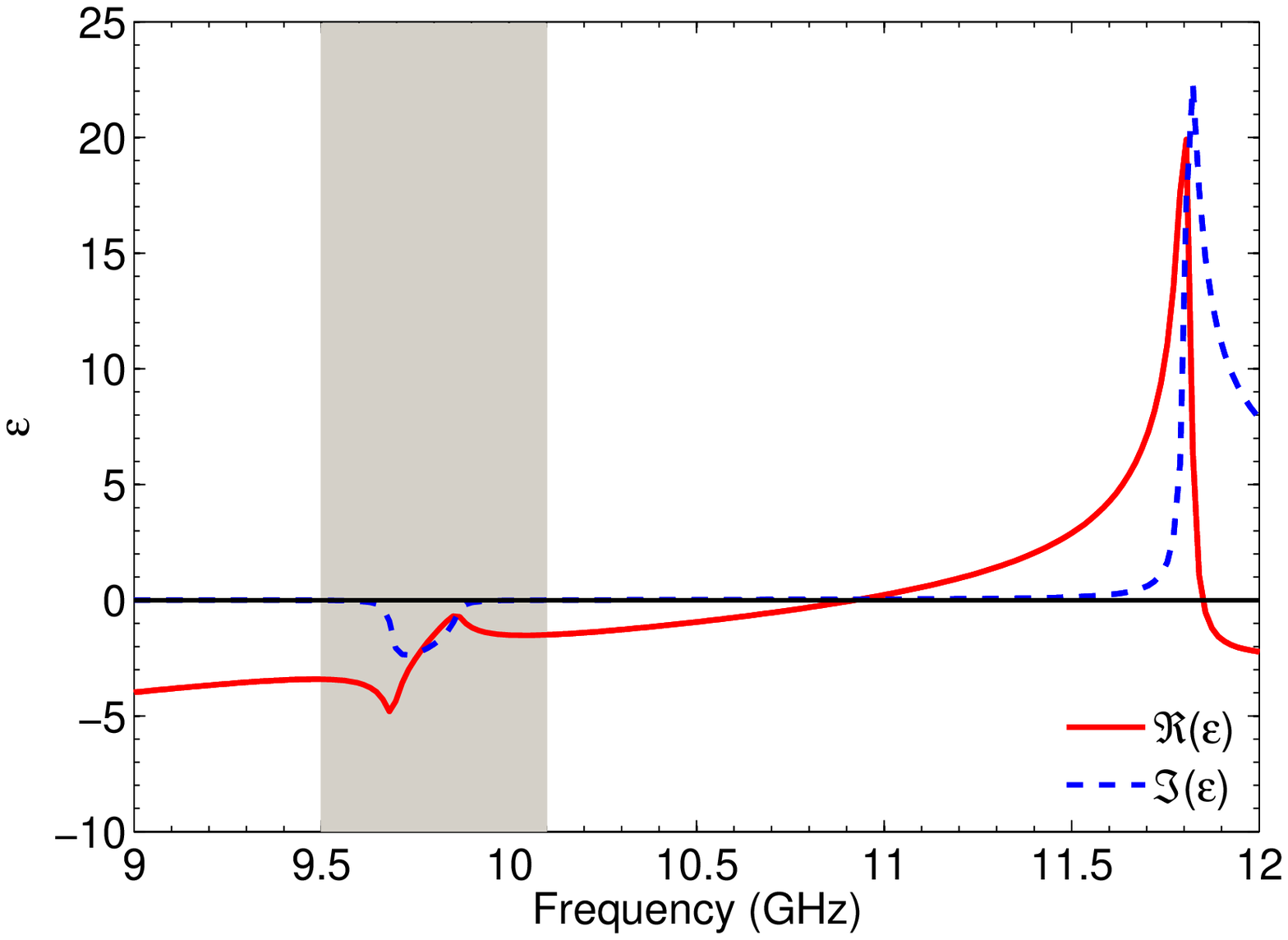}\\(c)\hspace*{3in}(d)\\[3pt]
\caption{(a) T, R and A versus frequencies for the complementary design.
(b) Retrieval results for the index of refraction, $n$. (c) the magnetic
permeability, $\mu$.  (d) and the electric permittivity, $\epsilon$.
Negative $\epsilon$ is obtained by the thin metallic surface and negative
$\mu$ is obtained by the complementary design. With single thin metallic surface
design, one obtains negative $n$.}
\label{fig6}
\end{figure*}

In \figref{fig3}a, we have a new metallic design that gives an electric dipole
and the complementary design gives a residual magnetic dipole shown in
\figref{fig3}b, for perpendicular direction on the metallic surface.

\section{New Designs for Babinet Principle}

In the previous section we reviewed the Babinet principle for different
designs (Figures \ref{fig1}--\ref{fig3}) and the complementary designs. These
designs (Figures \ref{fig1}--\ref{fig3}) can help us understand how the
electric dipole on the metallic film is related to he magnetic dipole of the
``complementary'' design. We can use these ideas of Babinet\ensuremath{'}s principle to use
metallic surface to give $\epsilon<0$ and $\mu<0$ and will have negative index
of refraction, $n<0$.

In \figref{fig4}a, we present a new metallic design that has no electric
dipole but a resonant electric quadrupole moment. Considering at first normal
incidence, the complementary structure show in \figref{fig4}b will possess a
resonant mode with a residual magnetic dipole moment perpendicular to the
plane as consequence of Babinet\ensuremath{'}s principle. We have thus established the
correspondence of a resonant electric quadrupole mode in the ``octopus''
structure to a resonant magnetic dipole mode in the complementary structure.

\red{Strictly, Babinet\ensuremath{'}s principle relates the field distributions between the direct structure and its complement when the wave propagates perpendicular to  the screen. However, here, we only use Babinet\ensuremath{'}s principle to establish the mapping of the resonant modes from the direct structure to its complement: If the direct structure has some resonant eigenmode, e.g. the electric cut-wire resonance of the finite-length wire, then we can establish using Babinet\ensuremath{'}s principle that also the complementary structure has a corresponding resonant eigenmode at this frequency but with all fields replaced by their dual fields. The resonant modes are a property of the geometry; it does not matter how the particular mode is excited. So, }the Babinet principle allows us to map fields between the corresponding resonances of mutually complementary structures. Therefore, after indentifying the resonances of the Babinet complement, we can elect to excite these resonances in a different geometry of incidence.

Now leaving the original Babinet setup behind, we realize that the
correspondence of the two resonant modes in the direct and complementary
structure, i.e., the derived ``translation rule'' of local charges, currents
and flux distributions on the surface, has to hold for any external
excitation. In order to couple to the magnetic dipole of the complementary
structure (\figref{fig4}b) we thus change the propagation direction to be
parallel with the surface such that the magnetic field of the incident wave
can couple to the ring current (i.e. the magnetic dipole) and we can obtain a
negative $\mu$. Notice the charges are shown in the ``legs'' of this design.
The current goes around and provides a magnetic dipole, which can give a
negative permeability. In addition the metallic surface (\figref{fig4}b) will
give negative permittivity, $\epsilon$. This complementary design can help us
to understand how to obtain negative $\mu$ and negative $\epsilon$. The design
shown in \figref{fig4} overlap with $\epsilon<0$ and $\mu<0$ and the impedance
$z=\sqrt{\mu/\epsilon}$ is very large and we have a lot of reflection and low
transmission. Deforming the structure while maintaining the same principle
topology of \figref{fig4}, we did a lot of simulations for different designs
to have very low impedance ($z=1$ for air) and to have very high transmission.
We find a new design (see \figref{fig5}) which satisfies the criterion
$\epsilon<0$ and $\mu<0$, and they have $|\epsilon|$ and $|\mu|$ the same
magnitude.

We will use an alternative design to obtain negative $\mu$ and the metallic
surface will give negative $\epsilon$ at the same frequency region, and the
magnitude of $\epsilon$ and $\mu$ will be comparable, and the impedance to be
close to 1. \red{The design is operating at 10GHz to keep sample fabrication and experimental verification simple but can be scaled to THz frequencies.} In \figref{fig5}a, we present a complementary design, which has a
big hole in the metallic surface that will give a negative $\mu$. This
structure is on a $35\mu m$ thick copper sheet with lattice constant
$a_x=a_y=10mm$ and $a_z=6mm$. In \figref{fig5}b, we plot the current density, which gives a
circular current and provides magnetic dipole momentum. We have used the
design of \figref{fig5}a, to obtain transmission, reflection, and absorption
$A=1-T-R$ with the CST. We simulated a periodic array of the proposed structures. With propagation direction parallel to the metallic
surface, we have used the amplitude and the phase of transmission and
reflection to obtain the frequency dependence of $\epsilon$, $\mu$, and $n$
through the retrieval procedure \cite{Smith2002}.

In \figref{fig6}, we plot the transmission, T, and reflection, R, and
absorption, A, versus the frequency. At the frequency 9.9 GHz, there is dip in
the reflection. In the retrieval results, we obtain negative $n$ with only one
metallic surface with the appropriate hole design. The hole design gives
negative $\mu$ and the metallic surface gives negative $\epsilon$, the
impedance is close to one, and the transmission is close to $65\%$. The figure of merit (FOM = $\mid$Re(n)/Im(n)$\mid$) is 9 with $n=-1.0$ with frequency 9.9GHz. The $FOM \simeq 8$ with $n=-0.5$ with frequency 9.95GHz. This is a
unique design that gives negative $\epsilon$ and $\mu$, with only one metallic
surface.

\section{Conclusion}

In conclusion, we have shown it is possible to construct Babinet-complement
structures (holes in planar metamaterials) that provide a resonant magnetic
moment perpendicular to the screen. We use such a magnetic resonance providing
$\mu<0$ together with the $\epsilon<0$ coming from the electric background
provided by the continuous regions of the screen, to obtain a double negative index
metamaterial for propagation direction parallel to the screen. In this
material, both $\epsilon<0$ and $\mu<0$ are provided by the same single
component. These designs can be used to fabricate three-dimensional
metamaterials and obtain negative index of refraction at microwave and
infrared wavelengths.

\bigskip\noindent\emph{Acknowledgement:} Work at Ames Laboratory was
partially supported by the Department of Energy (Basic Energy Sciences,
Division of Materials Sciences and Engineering) under Contract No.\
DE-AC02-07CH11358 (computational studies) \red{and by ERC grant No. 320081 (PHOTOMETA).} This work was partially supported
by the Office of Naval Research, Award No.\ N00014-10-1-0925.

\bigskip\noindent
$*$ email: \texttt{mywaters@iastate.edu}\\
$\dag$ email: \texttt{soukoulis@ameslab.gov}
\vfill

\bibliographystyle{apsrev}

\end{document}